\documentclass[%
 reprint,
superscriptaddress,
 amsmath,amssymb,
 aps,
prb,
]{revtex4-2}

\usepackage{graphicx}
\usepackage{dcolumn}
\usepackage{bm}
\usepackage[colorlinks,linkcolor=blue,citecolor=blue,urlcolor=black,hyperindex,driverfallback=dvipdfm]{hyperref}

\newcommand{\plasmon}{\mathrm{pl}} 
\newcommand{\imag}{{i\mkern1mu}} 
\newcommand{\eff}{\mathrm{eff}} 
\newcommand{\sigmasca}{\sigma_{\mathrm{sca}}} 
\newcommand{\eqcomma}{\,\mathrm{,}} 
\newcommand{\eqdot}{\,\mathrm{.}} 
\usepackage{xcolor}
\usepackage{upgreek}

\renewcommand{\Im}{\operatorname{Im}}

\begin{document}

\title{Strong coupling in a Au plasmonic antenna--SiO$_{2}$ layer system: a hybrid mode analysis}

\author{Pavel Gallina}
\email{pavel.gallina@ceitec.vutbr.cz}
\affiliation{CEITEC BUT, Brno University of Technology, Technická 10, 616 00 Brno, Czech Republic}
\affiliation{Institute of Physical Engineering, Brno University of Technology, Technická 2, 616 69 Brno, Czech Republic}
\author{Michal Kvapil}
\affiliation{CEITEC BUT, Brno University of Technology, Technická 10, 616 00 Brno, Czech Republic}
\affiliation{Institute of Physical Engineering, Brno University of Technology, Technická 2, 616 69 Brno, Czech Republic}
\author{Ji\v{r}\'{i} Li\v{s}ka}
\affiliation{CEITEC BUT, Brno University of Technology, Technická 10, 616 00 Brno, Czech Republic}
\author{Andrea Kone\v{c}n\'{a}}
\affiliation{CEITEC BUT, Brno University of Technology, Technická 10, 616 00 Brno, Czech Republic}
\author{\mbox{Vlastimil K\v{r}\'{a}pek}}
\affiliation{CEITEC BUT, Brno University of Technology, Technická 10, 616 00 Brno, Czech Republic}
\affiliation{Institute of Physical Engineering, Brno University of Technology, Technická 2, 616 69 Brno, Czech Republic}
\author{Radek Kalousek}
\affiliation{CEITEC BUT, Brno University of Technology, Technická 10, 616 00 Brno, Czech Republic}
\affiliation{Institute of Physical Engineering, Brno University of Technology, Technická 2, 616 69 Brno, Czech Republic}
\author{Jakub Zl\'{a}mal}
\affiliation{CEITEC BUT, Brno University of Technology, Technická 10, 616 00 Brno, Czech Republic}
\affiliation{Institute of Physical Engineering, Brno University of Technology, Technická 2, 616 69 Brno, Czech Republic}
\author{Tom\'{a}\v{s} \v{S}ikola}
\affiliation{CEITEC BUT, Brno University of Technology, Technická 10, 616 00 Brno, Czech Republic}
\affiliation{Institute of Physical Engineering, Brno University of Technology, Technická 2, 616 69 Brno, Czech Republic}

\begin{abstract}
A detailed analysis of the optical response of a system accommodating several coupled modes is needed for the complete understanding of the strong coupling effect. In this paper, we report on the analysis of scattering cross section spectra of Au antennas on a SiO$_{2}$ layer on a Si substrate in the IR region. A classical model of coupled oscillators is used for determining the resonant energies, damping rates and coupling strengths of four phonon polariton modes in the SiO$_{2}$ layer coupled to a localized surface plasmon mode in a Au antenna. The calculated Hopfield mixing coefficients then show the contribution of the individual uncoupled modes to the hybrid modes of the coupled system.
\end{abstract}

\maketitle

\section{Introduction}
Plasmonic antennas have been widely used for enhancing an optical response of materials or for fabricating metamaterials with new exotic properties. In both cases, localized surface plasmon (LSP) resonances in the antennas are utilized~\cite{spektor2015}. However, at specific properties of the substrate or surrounding medium, one can observe a completely different behaviour of these antennas caused by a coupling of plasmons with other excitations, for instance phonons or excitons. The so called weak coupling regime is characterized only by minor perturbations of the shape of measured spectra and it mainly contributes to an electric near-field enhancement resulting in a stronger photon absorption and emission of the materials~\cite{AlonsoGonzlez2012,moskovits2005}. On the other hand, the strong coupling leads to a profound shape change of antenna response spectra as new hybrid resonance modes are created in the system. One can particularly observe the Rabi splitting of the resonance peaks (and the anticrossing in dispersion relations of the modes) and eventual opening of a transparency window, as well as changes in excitation lifetimes~\cite{dintinger2005}.

Surface (interface) phonon polaritons (SPhP, IPhP) are excitations arising from the coupling of photons and phonons at dielectric material surfaces (interfaces), similarly to surface plasmon polaritons in conducting materials~\cite{caldwell2015,yu2010}. The transverse optical (TO) phonons can be excited directly by light, however, their excitation efficiency in a thin layer is relatively weak (in comparison with plasmons in antennas). The longitudinal optical (LO) phonons as well as surface phonon polaritons cannot be excited directly by plane waves at normal incidence on a planar sample surface. The electric near field of plasmonic particles deposited or fabricated on a surface supporting phononic excitations may not only act in enhancing the light absorption or excitation of modes that cannot couple directly with light, but the associated localized surface plasmons can also couple with the phonons and phonon polaritons.

We present an analysis of the electromagnetic coupling between localized surface plasmons in gold rectangular antennas and phonon polaritons in a silicon dioxide thin film on a semi-infinite silicon substrate (see Fig.~\ref{fig_scheme_dielf_disprel}a). Even though the underlying principles in this system have been previously described~\cite{huck2016,shelton2011}, a detailed analysis of the resultant hybrid modes based on a coupled-oscillator model, realistic SiO$_{2}$ dielectric function, as well as on the calculation of the Hopfield mixing coefficients has not been performed.

\section{Results}

\begin{figure*}
\includegraphics[width=\textwidth]{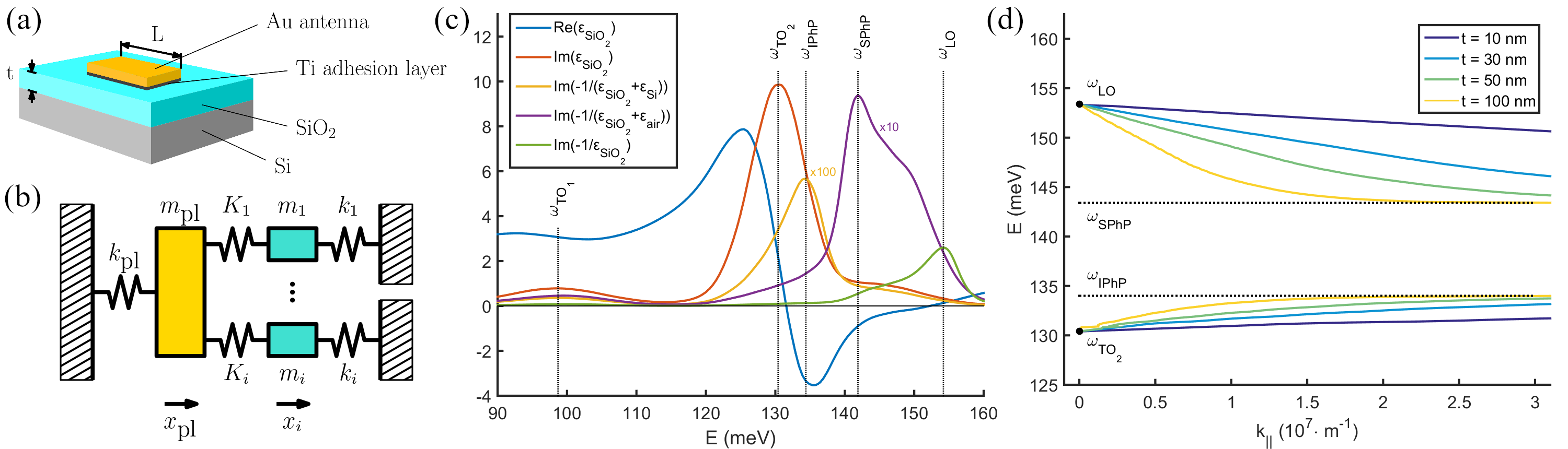}
\caption{(a) Schematic of the investigated system. (b) Schematic of the coupled-oscillator model. (c) Relative dielectric function and functions proportional to the energy loss of the system (maxima positions correspond to bulk and surface phonon modes). (d) Dispersion relation of phonon polaritons for different thicknesses of the SiO$_{2}$ layer on the Si substrate.}
\label{fig_scheme_dielf_disprel}
\end{figure*}

The investigation of the strong coupling effect between localized surface plasmons in metallic antennas and phonon polaritons in an absorbing dielectric layer was performed on a system depicted in Fig.~\ref{fig_scheme_dielf_disprel}a. It consists of a Au rectangular antenna of a variable length $L$ (between 0.8 and 3.6 $\upmu$m), width of 400~nm, and height of 25~nm placed on top of a SiO$_{2}$ layer of a thickness $t$ (varying from 5 to 200~nm) and deposited on a Si substrate. A 5~nm-thick Ti adhesion layer between the antenna and SiO$_{2}$ was used. The samples were fabricated by atomic layer deposition of SiO$_{2}$ on a double side polished Si substrate, and a subsequent electron-beam lithography process followed by an electron-beam evaporation of Ti and Au. Reflectance and trasmittance spectra were measured by Fourier-transform infrared micro-spectroscopy. Finite-difference time-domain simulations (using the Lumerical software~\cite{lumerical}) for the antenna configuration and dimensions identical with those of experimental antennas were then performed. There is a good agreement between the experimental and simulated reflectance and transmittance spectra (see Fig.~\textcolor{blue}{S1} in Supplemental Material), both showing a similar peak splitting and opening of the transparency window around an energy of 150~meV. It enables us to preferentially use the smoother simulated spectra rather than the noisy experimental ones for fitting purposes discussed further. Moreover, the scattering cross section spectra to be fitted by curves derived from an analytical model could be acquired only from the simulations.

The realistic dielectric functions used in simulations were taken from Palik~\cite{palik1998} for gold, titanium and silicon and from Kischkat~\cite{kischkat2012} for silicon dioxide. The resonant energy of the dipolar mode of the localized surface plasmon in the antenna approximately depends linearly on the inverse value of the antenna length~$L$, and can be thus swept over the whole investigated spectral range (80-180~meV) by changing the length~$L$. There is also a higher LSP mode for longer antennas on a thinner silicon dioxide layer, however, the energy range in our analysis was limited to values below the energy of this mode so that this mode would not affect the fitting process. A multitude of phonon polariton modes are present in the silicon dioxide layer on the silicon substrate, which can be separated into two regions corresponding to the Si-O-Si stretching vibrations~\cite{gunde2000}. The silicon dioxide dielectric function and the relevant energy-loss functions are depicted in Fig.~\ref{fig_scheme_dielf_disprel}c. The asymmetric stretching vibration mode is described by an oscillator with the TO phonon energy $\hbar\omega_{\mathrm{TO_{2}}}=130.4$~meV (1052~cm$^{-1}$, 9.5~$\upmu$m), which corresponds to the maximum of $\Im\left(\varepsilon_{\mathrm{SiO_{2}}}\right)$. The energy of the LO bulk phonon $\hbar\omega_{\mathrm{LO}}=154.2$~meV (1244~cm$^{-1}$, 8.0~$\upmu$m) is defined by the maximum of an energy-loss function $\Im\left(-1/\varepsilon_{\mathrm{SiO_{2}}}\right)$~\cite{luth2001}. The energy of a surface phonon polariton (SPhP) $\hbar\omega_{\mathrm{SPhP}}=141.9$~meV (1146~cm$^{-1}$, 8.7~$\upmu$m) appears at the maximum of the energy-loss function $\Im\left(-1/\left(\varepsilon_{\mathrm{SiO_{2}}}+\varepsilon_{\mathrm{air}}\right)\right)$, and the energy of an interface phonon polariton (IPhP) $\hbar\omega_{\mathrm{IPhP}}=134.4$~meV (1084~cm$^{-1}$, 9.2~$\upmu$m) appears at the maximum of the energy-loss function $\Im\left(-1/\left(\varepsilon_{\mathrm{SiO_{2}}}+\varepsilon_{\mathrm{Si}}\right)\right)$. The actual phonon polaritons that couple with the LSP are located either on the LO-SPhP or the TO-IPhP branch in the phonon polariton dispersion relation shown in Fig.~\ref{fig_scheme_dielf_disprel}d. The dispersion relation was numerically calculated using the formula in Eq.~(\textcolor{blue}{S1}) (in Supplemental Material) with a reduced damping. A symmetrical stretching vibration mode is also present at lower energies with a TO phonon energy $\hbar\omega_{\mathrm{TO_{1}}}=98.7$~meV (796~cm$^{-1}$, 12.6~$\upmu$m).

In order to obtain the parameters defining the coupled system, a classical model of coupled oscillators has been employed~\cite{novotny2010}. We start with equations of motion of five damped harmonic oscillators, where four of these oscillators (corresponding to phonon polaritons, marked by the indices 1-4) couple with the fifth one (localized surface plasmon, marked by the index $pl$) but not with each other (Fig.~\ref{fig_scheme_dielf_disprel}b). Moreover, we consider that an external driving force $F_{\plasmon}$ only acts on the fifth oscillator (LSP), since the direct excitation of phonon polaritons by light is negligible. The equations thus take the form:

\begin{widetext}
\begin{subequations}
\begin{alignat}{8}
&m_{1}\ddot{x}_{1}&&+b_{1}\dot{x}_{1}&&+k_{1}x_{1}&&+K_{1}(x_{1}-x_{\plasmon})&& && && &&=0\eqcomma\label{harm_osc_gen_nov_1}\\
&m_{2}\ddot{x}_{2}&&+b_{2}\dot{x}_{2}&&+k_{2}x_{2}&& &&+K_{2}(x_{2}-x_{\plasmon})&& && &&=0\eqcomma\label{harm_osc_gen_nov_2}\\
&m_{3}\ddot{x}_{3}&&+b_{3}\dot{x}_{3}&&+k_{3}x_{3}&& && &&+K_{3}(x_{3}-x_{\plasmon})&& &&=0\eqcomma\label{harm_osc_gen_nov_3}\\
&m_{4}\ddot{x}_{4}&&+b_{4}\dot{x}_{4}&&+k_{4}x_{4}&& && && &&+K_{4}(x_{4}-x_{\plasmon})&&=0\eqcomma\label{harm_osc_gen_nov_4}\\
&m_{\plasmon}\ddot{x}_{\plasmon}&&+b_{\plasmon}\dot{x}_{\plasmon}&&+k_{\plasmon}x_{\plasmon}&&-K_{1}(x_{1}-x_{\plasmon})&&-K_{2}(x_{2}-x_{\plasmon})&&-K_{3}(x_{3}-x_{\plasmon})&&-K_{4}(x_{4}-x_{\plasmon})&&=F_{\plasmon}\eqcomma\label{harm_osc_gen_nov_pl}
\end{alignat}
\label{harm_osc_gen_nov}
\end{subequations}
\end{widetext}

\noindent where $x_{i/\plasmon}$, $m_{i/\plasmon}$, $b_{i/\plasmon}$ and $k_{i/\plasmon}$ correspond to displacement, mass, damping constant and restoring force constant, respectively, for each oscillator. $K_{i}$ is the constant corresponding to the coupling of the $i$-th phonon polariton with the LSP. We then define terms:

\begin{subequations}
\begin{equation}
\gamma_{i/\plasmon}=\frac{b_{i/\plasmon}}{2m_{i/\plasmon}}\eqcomma
\end{equation}
\begin{equation}
\omega_{i}=\sqrt{\frac{k_{i}+K_{i}}{m_{i}}}\eqcomma
\label{def_omega_i}
\end{equation}
\begin{equation}
\omega_{\plasmon}=\sqrt{\frac{k_{\plasmon}+K_{1}+K_{2}+K_{3}}{m_{\plasmon}}}\eqcomma
\label{def_omega_plasmon}
\end{equation}
\begin{equation}
F_{\plasmon}=Eq_{\plasmon}\eqcomma
\end{equation}
\end{subequations}

\noindent with $\gamma_{i/\plasmon}$ and $\omega_{i/\plasmon}$ being the damping rates and resonant energies of uncoupled oscillators, $E$ the electric field of the incident light, and $q_{\plasmon}$ the effective charge representing the LSP. Furthermore, as the electric field $E$ has the form of a harmonic plane wave with frequency $\omega$, all the displacements $x_{i/\plasmon}$ are time dependent according to $e^{-i\omega t}$.

If there was no coupling present ($K_{i}=0$), the individual uncoupled oscillators would be characterised by a term proportional to their polarizability (Lorentzian oscillator)

\begin{equation}
\alpha'_{i/\plasmon}=\frac{1}{\omega_{i/\plasmon}^{2}-\omega^2-2\imag\gamma_{i/\plasmon}\omega}\eqdot
\label{harm_osc_alpha}
\end{equation}

The coupling can be described by the coupling strength constants~\cite{novotny2010}

\begin{equation}
g_{i}=\frac{\sqrt{K_{i}/m_{i}}\sqrt{K_{i}/m_{\plasmon}}}{2\sqrt{\omega_{i}\omega_{\plasmon}}}\eqcomma
\label{def_g}
\end{equation}

\noindent which agree with the coupling strength constants from a Jaynes-Cummings quantum-mechanical model~\cite{casanova2010}, and $2g_{i}$ is the frequency splitting at the anticrossing which is characteristic for the strong coupling~\cite{novotny2010,torma2014}.

By solving Eqs.~\eqref{harm_osc_gen_nov} and comparing the solution with the formula for polarizability $\alpha=\frac{q_{\plasmon}x_{\plasmon}}{E}$, we get the expression for the polarizability of the system in the form

\begin{equation}
\alpha=\frac{\alpha'_{\plasmon}\varphi_{\plasmon}}{1 - \sum\limits_{i=1}^{4} 4g_{i}^{2}\omega_{i}\omega_{\plasmon}\alpha'_{i}\alpha'_{\plasmon}}\eqcomma
\label{alpha_gen_nov}
\end{equation}

\noindent where $\varphi_{\plasmon}=\frac{q_{\plasmon}^{2}}{m_{\plasmon}}$ is the constant describing the oscillator representing the LSP and denoting the efficiency of the system excitation with the external field. Since the scattering particle supporting the localized surface plasmon has a finite volume, an effective radiation-corrected polarizability has to be considered~\cite{jackson1999}:

\begin{equation}
\alpha_{\eff}=\frac{\alpha}{1-\imag\frac{1}{6\pi\varepsilon_{0}}\left(\frac{\omega}{c}\right)^{3}\alpha}\eqcomma
\label{alpha_rc_nov}
\end{equation}

\noindent where $c$ is the speed of light in vacuum, and $\varepsilon_{0}$ is the vacuum permittivity. The expression for the scattering cross section which was used for fitting the simulation results can be then calculated as~\cite{novotny2012}:

\begin{equation}
\sigmasca=\frac{1}{6\pi\varepsilon_{0}^{2}}\left(\frac{\omega}{c}\right)^{4} \vert\alpha_{\eff}\vert^2\eqdot
\label{csca_def_nov}
\end{equation}

\begin{figure}[b]
\includegraphics[width=0.5\textwidth]{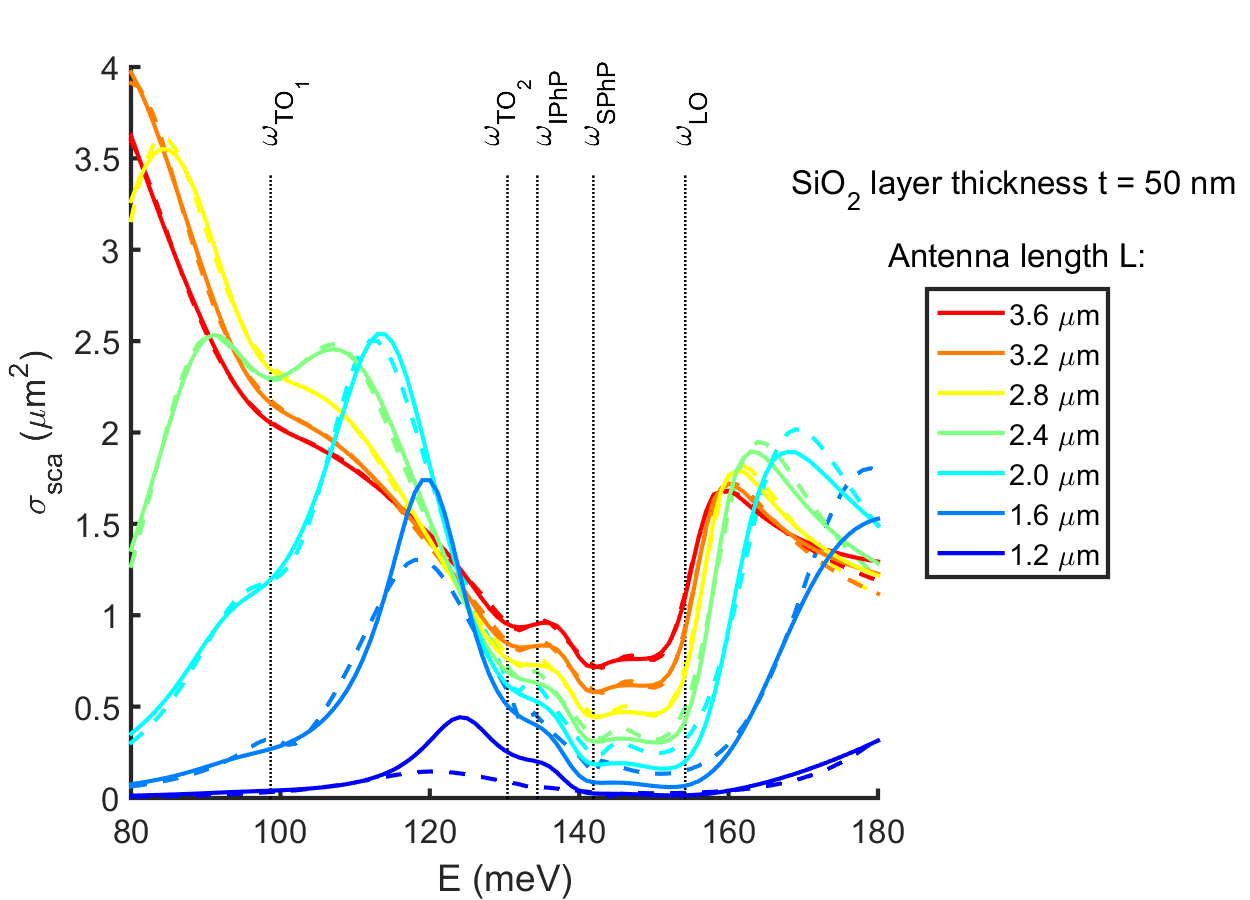}
\caption{Calculated (solid lines) and fitted (dashed lines) scattering cross section spectra of gold antennas for different lengths $L$ on the SiO$_{2}$ layer with the thickness $t=50$~nm on the Si substrate. Splitting is visible around the resonant frequencies of SiO$_{2}$ phonon polariton modes (vertical lines).}
\label{fig_sigma_sca_t}
\end{figure}

\begin{figure*}
\includegraphics[width=\textwidth]{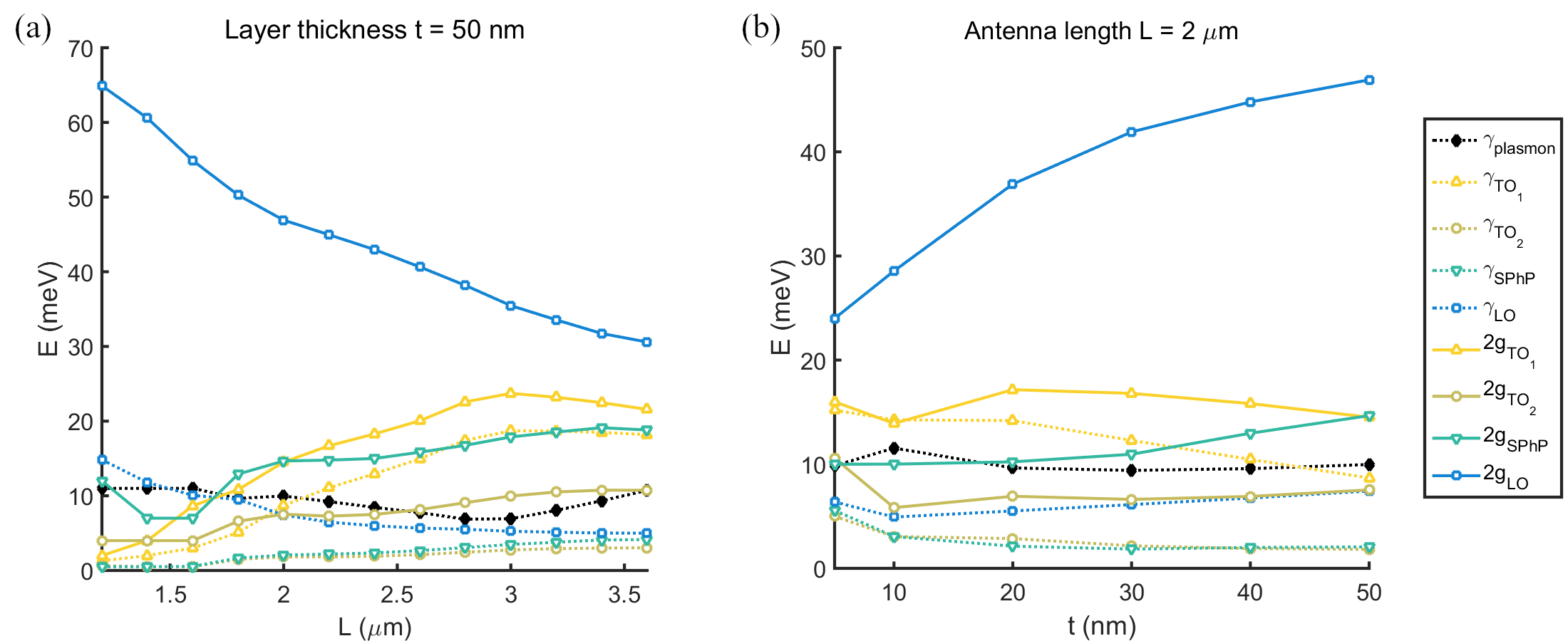}
\caption{Dependence of the damping rates and coupling strengths (a) on the antenna length $L$ for the constant SiO$_{2}$ layer thickness $t=50$~nm, and (b) on the layer thickness $t$ for the constant antenna length $L=2~\upmu$m.}
\label{fig_g_gamma}
\end{figure*}

Simulated scattering cross section spectra $\sigmasca$ for a constant thickness of the SiO$_{2}$ layer ($t=50$~nm) and different antenna lengths $L$ are shown in Fig.~\ref{fig_sigma_sca_t} (solid lines). The spectra for the constant length $L=2~\upmu$m and different thickness $t$ are in Fig.~\textcolor{blue}{S2} in Supplemental Material. With the increasing antenna length $L$ the plasmon resonance redshifts. However, the peak splitting (minima) is in all spectra at almost the same energies, which corresponds to the coupling of the phonon polariton modes with the plasmon resonance mode. By fitting the spectra with Eq.~\eqref{csca_def_nov} (dashed lines in Fig.~\ref{fig_sigma_sca_t}), we obtain the resonant energies $\omega_{i/\plasmon}$ and damping rates $\gamma_{i/\plasmon}$ of the original localized surface plasmon and phonon polaritons (which can be then assigned to the individual phonons), and coupling strengths $g_{i}$. The coupling phonon polariton modes are at energies around 101~meV (TO$_{1}$), 132~meV (TO$_{2}$), 144~meV (SPhP) and 152~meV (LO). Only four oscillators for phonons were used during the fitting, since the addition of another oscillator for IPhP made a negligible impact, pointing to the conclusion that either the TO$_{2}$ phonon excitation is much stronger and thus the effect of coupling of the plasmon with the IPhP mode is unobservable~\cite{huck2016}, or that all the TO$_{2}$-IPhP branch mode energies are so close to each other that it can be described by a single oscillator during the fitting process.

The dependence of the coupling strengths $g_{i}$ and damping rates $\gamma_{i}$ 
on the Au antenna length~$L$ or the SiO$_{2}$ layer thickness~$t$ is shown in Fig.~\ref{fig_g_gamma}. The condition for the strong coupling stated as $2g_{i}>\gamma_{i}+\gamma_{\plasmon}$~\cite{torma2014,autore2018} is fulfilled for the LO phonon mode, with the highest coupling strength corresponding to the splitting of $2g_{\mathrm{LO}}$ around 50-60~meV for smaller~$L$ and larger~$t$. The coupling of the other modes with the LSP (with $2g_{\mathrm{TO_{1}}}$ around 20~meV, $2g_{\mathrm{TO_{2}}}$ around 8~meV, and $2g_{\mathrm{SPhP}}$ around 20~meV) is also observable in the scattering cross section spectra, however, the values of the coupling strengths are not particularly larger than the damping rates, therefore there is just an onset of the strong coupling regime for these modes.

\begin{figure}[t]
\includegraphics[width=0.5\textwidth]{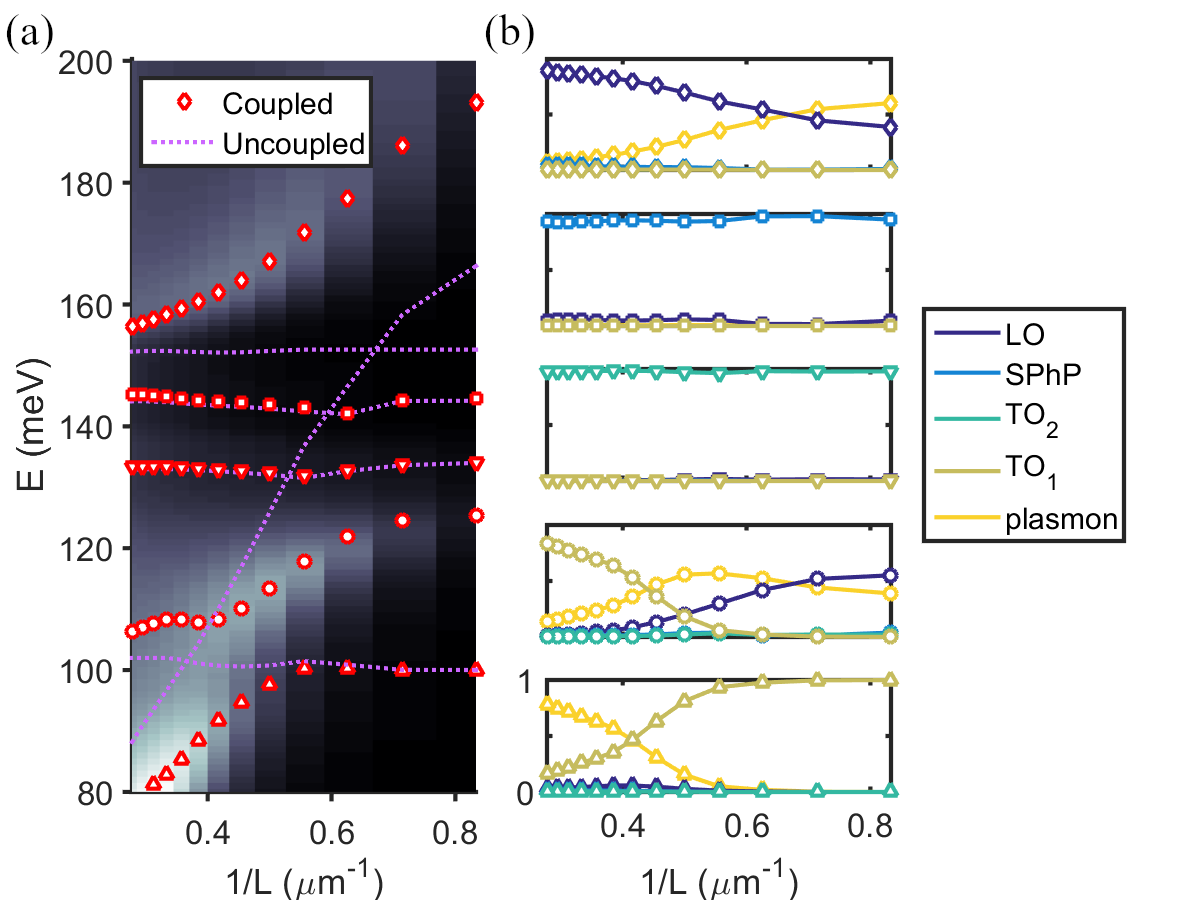}
\caption{(a) Dependence of the simulated scattering cross section spectra (background), uncoupled mode energies from fitting (magenta lines) and hybrid mode energies from subsequent analysis (red marks) on the inverse values of the antenna length $L$ for the SiO$_{2}$ layer thickness $t=50$~nm. (b) Hopfield mixing coefficients showing relative contributions of uncoupled modes to the new hybrid modes. The bottom graph corresponds to the bottom branch in the image on the left (similarly for the other graphs and branches), the values for the Hopfield mixing coefficients are between 0 and 1.}
\label{fig_hopfield}
\end{figure}

The parameters obtained from fitting can be further utilized for an analysis of the hybrid modes given by the coupling~\cite{brinek2018,slootsky2014,zhang2021}. A simplified matrix describing the system

\begin{equation}
\bm{H} = 
\begin{pmatrix}
\omega_{1} & 0 & 0 & 0 & g_{1}\\
0 & \omega_{2} & 0 & 0 & g_{2}\\
0 & 0 & \omega_{3} & 0 & g_{3}\\
0 & 0 & 0 & \omega_{4} & g_{4}\\
g_{1} & g_{2} & g_{3} & g_{4} & \omega_{\plasmon}
\end{pmatrix}
\end{equation}

\noindent can be constructed either directly from the Jaynes-Cummings quantum mechanical model or from the model presented here by assuming that all the energies (both $\omega_{i/\plasmon}$ and $\omega$) are close to each other~\cite{autore2018} and omitting the damping. The eigenvalues of the matrix correspond to the energies of the new hybrid modes (peak positions in the spectra). The eigenvectors $(X_{1},X_{2},X_{3},X_{4},X_{\plasmon})_{j}^{T}$, where $j=1,2,3,4,5$ for the five eigenvalues, can be further utilized to find the relative contribution of individual uncoupled plasmon or phonon polariton modes into the hybrid modes. The absolute squares of the eigenvector components ($|X_{1}|^{2}$, $|X_{2}|^{2}$, $|X_{3}|^{2}$, $|X_{4}|^{2}$ and $|X_{\plasmon}|^{2}$) give the Hopfield mixing coefficients which express the fractional contribution of the individual modes to the new hybrid mode~\cite{brinek2018,slootsky2014,zhang2021}. Fig.~\ref{fig_hopfield}a shows the peak positions (eigenvalues of the matrix $\bm{H}$, red points) laid over the plot given by the simulated scattering cross section $\sigmasca$ (in the background), together with the uncoupled mode energies $\omega_{i/\plasmon}$ from fitting (dotted lines). The plots of the Hopfield mixing coefficients in Fig.~\ref{fig_hopfield}b are linked to the corresponding branches in Fig.~\ref{fig_hopfield}a. They show a gradual shift of the major contributor to the two lowest and the one highest energy coupled modes (e.g. the lowest branch being mainly plasmon-like for longer antennas and phonon-like for shorter antennas) with $1/L$. Interestingly, the two branches around 132 and 144~meV seem to be composed only from TO$_{2}$ and SPhP modes, respectively, even though they are coupled with the plasmons as there is a visible minimum (splitting) in the scattering cross section spectra and the coupling strengths $g_{i}$ have a comparable magnitude to that of the TO$_{1}$ mode. This could be possibly explained by the LO phonon being a dominant mode producing the whole splitting (transparency window) between 120-180~meV and TO$_{2}$ and SPhP having only a small effect and being described rather as only weakly coupled. Contrarily, the TO$_{1}$ mode is at the onset of strong coupling, since it is not affected by the LO mode (which is far away) and produces anticrossing behaviour around 100~meV, even though having a similar coupling strength and damping rate as TO$_{2}$ or SPhP modes. Moreover, for a thinner layer the splitting caused by the LO mode is smaller, and the splitting (anticrossing) around the TO$_{2}$ mode starts to be more noticeable together with a slight contribution of the TO$_{2}$ mode to the lower energy coupled mode, as shown in Fig.~\textcolor{blue}{S3}.

\section{Conclusions}
We have presented the method for analysis of the optical response of a coupled system of localized surface plasmons and phonon polaritons by utilizing the classical coupled-oscillator model and the Hopfield mixing coefficients, providing a better understanding of the strong coupling effect. The fitting of the scattering cross section spectra by the model revealed coupling of LSPs in the Au antenna with several modes in the SiO$_{2}$ layer. More specifically, the onset of the strong coupling around an energy of 100~meV, observable as the peak splitting, can be described as an interaction of a single phonon with a plasmon mode. However, the spectral features between 120 and 180 meV are caused by tranverse and longitudinal optical phonons and surface phonon polaritons, of which the strong coupling of the LO mode (with a peak splitting of up to 60~meV) is the most prominent interaction effect, influencing also the coupling of the other modes. Even though there should also be an interface phonon polariton mode, it has not been used in the fitting of the spectra due to its negligible effect. The calculation of the Hopfield mixing coefficients describing the contribution of the uncoupled modes to the hybrid modes further supports these results. There is a major contribution of the LO mode to the hybrid modes surrounding the transparency window, while the TO$_{2}$ and SPhP modes seem not to mix with the other modes despite the minor splitting in the spectra, hinting to only a weak coupling regime of those two modes.

\begin{acknowledgments}
We acknowledge the support by the Czech Science Foundation (Grant No.*20-28573S*), European Commission (H2020-Twininning project No. 810626 – SINNCE, M-ERA NET HYSUCAP/TACR-TH71020004),*BUT* – specific research No.*FSI-S-20-648*5, and Ministry of Education, Youth and Sports of the Czech Republic (CzechNanoLab Research Infrastructure – LM2018110).
\end{acknowledgments}

\bibliography{Strong_coupling_in_a_Au_plasmonic_antenna_SiO2_layer_system_a_hybrid_mode_analysis}

\end{document}


\title{Strong coupling in a Au plasmonic antenna--SiO$_{2}$ layer system: a hybrid mode analysis\\
Supplemental Material}%

\author{Pavel Gallina}
\affiliation{CEITEC BUT, Brno University of Technology, Technická 10, 616 00 Brno, Czech Republic}
\affiliation{Institute of Physical Engineering, Brno University of Technology, Technická 2, 616 69 Brno, Czech Republic}
\author{Michal Kvapil}
\affiliation{CEITEC BUT, Brno University of Technology, Technická 10, 616 00 Brno, Czech Republic}
\affiliation{Institute of Physical Engineering, Brno University of Technology, Technická 2, 616 69 Brno, Czech Republic}
\author{Ji\v{r}\'{i} Li\v{s}ka}
\affiliation{CEITEC BUT, Brno University of Technology, Technická 10, 616 00 Brno, Czech Republic}
\author{Andrea Kone\v{c}n\'{a}}
\affiliation{CEITEC BUT, Brno University of Technology, Technická 10, 616 00 Brno, Czech Republic}
\author{\mbox{Vlastimil K\v{r}\'{a}pek}}
\affiliation{CEITEC BUT, Brno University of Technology, Technická 10, 616 00 Brno, Czech Republic}
\affiliation{Institute of Physical Engineering, Brno University of Technology, Technická 2, 616 69 Brno, Czech Republic}
\author{Radek Kalousek}
\affiliation{CEITEC BUT, Brno University of Technology, Technická 10, 616 00 Brno, Czech Republic}
\affiliation{Institute of Physical Engineering, Brno University of Technology, Technická 2, 616 69 Brno, Czech Republic}
\author{Jakub Zl\'{a}mal}
\affiliation{CEITEC BUT, Brno University of Technology, Technická 10, 616 00 Brno, Czech Republic}
\affiliation{Institute of Physical Engineering, Brno University of Technology, Technická 2, 616 69 Brno, Czech Republic}
\author{Tom\'{a}\v{s} \v{S}ikola}
\affiliation{CEITEC BUT, Brno University of Technology, Technická 10, 616 00 Brno, Czech Republic}
\affiliation{Institute of Physical Engineering, Brno University of Technology, Technická 2, 616 69 Brno, Czech Republic}

\maketitle

The dispersion relation in Fig.~\textcolor{blue}{1}d was obtained numerically by solving the formula

\begin{equation}
e^{\sqrt{k_{\parallel}^{2}-\varepsilon_{2}\frac{\omega^{2}}{c^{2}}}} = \frac{\frac{\sqrt{k_{\parallel}^{2}-\varepsilon_{2}\frac{\omega^{2}}{c^{2}}}}{\varepsilon_{2}}-\frac{\sqrt{k_{\parallel}^{2}-\varepsilon_{1}\frac{\omega^{2}}{c^{2}}}}{\varepsilon_{1}}}{\frac{\sqrt{k_{\parallel}^{2}-\varepsilon_{2}\frac{\omega^{2}}{c^{2}}}}{\varepsilon_{2}}+\frac{\sqrt{k_{\parallel}^{2}-\varepsilon_{1}\frac{\omega^{2}}{c^{2}}}}{\varepsilon_{1}}}\frac{\frac{\sqrt{k_{\parallel}^{2}-\varepsilon_{2}\frac{\omega^{2}}{c^{2}}}}{\varepsilon_{2}}-\frac{\sqrt{k_{\parallel}^{2}-\varepsilon_{3}\frac{\omega^{2}}{c^{2}}}}{\varepsilon_{3}}}{\frac{\sqrt{k_{\parallel}^{2}-\varepsilon_{2}\frac{\omega^{2}}{c^{2}}}}{\varepsilon_{2}}+\frac{\sqrt{k_{\parallel}^{2}-\varepsilon_{3}\frac{\omega^{2}}{c^{2}}}}{\varepsilon_{3}}}\eqcomma
\label{dispersion_relation}
\end{equation}

\noindent where $c$ is the speed of light in vacuum, and $\varepsilon_{i}$ are the relative dielectric functions of the layers, i.e. air (vacuum), SiO$_{2}$ and Si. A realistic dielectric function with reduced damping was used, since the damping would prevent obtaining the values for higher~$k_{\parallel}$. This was achieved by fitting the dielectric function with a single Lorentzian oscillator as $\varepsilon=\varepsilon_{\infty} + \frac{\varepsilon_{\mathrm{L}}}{\omega_{0}^{2}-\omega^{2}-2\imag\omega\gamma}$ in order to find $\varepsilon_{\infty}=1.96$, and then reducing the imaginary part of the inverse term of $\left(\varepsilon-\varepsilon_{\infty}\right)$ to 1~\% of its value, reaching an approximate reduction of the damping rate ($\gamma$ in the Lorentzian oscillator formula) while keeping the complex features of the dielectric function to some degree. This process led to a slight shift of the resonant energies, nonetheless, this is negligible in comparison to the minor differences obtained during the fitting of $\sigmasca$ spectra.

\begin{figure}[h]
\includegraphics[width=\textwidth]{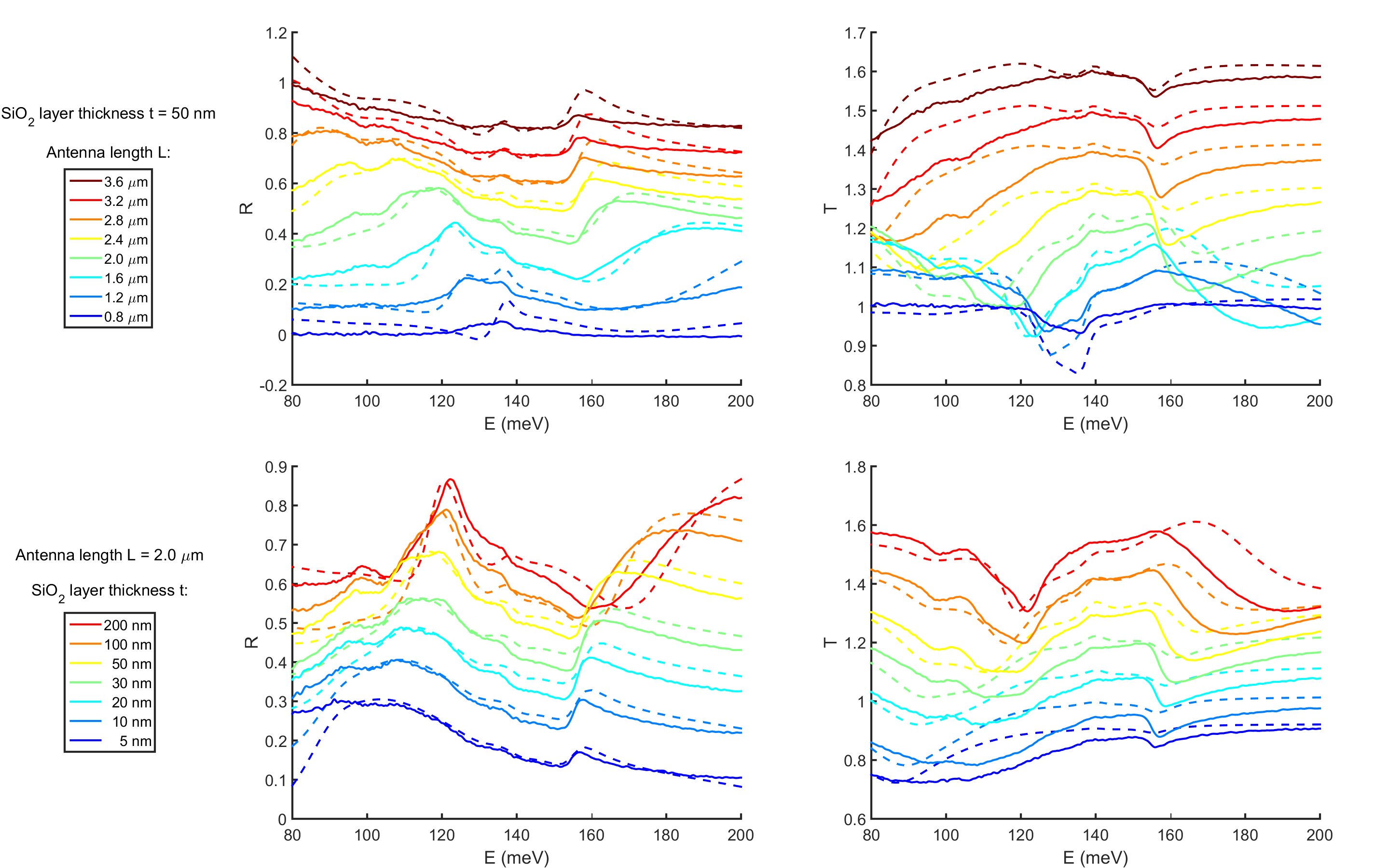}
\caption{Comparison of reflectance (left) and transmittance (right) spectra obtained from FTIR measurements (solid lines) with those provided by FDTD simulations (dashed lines) for different antenna lengths (top) and different SiO$_{2}$ layer thicknesses (bottom). The values correspond to the blue lines, with the other lines being shifted.}
\label{fig_experiment_simulation_RT}
\end{figure}

\begin{figure}[h]
\includegraphics[width=0.5\textwidth]{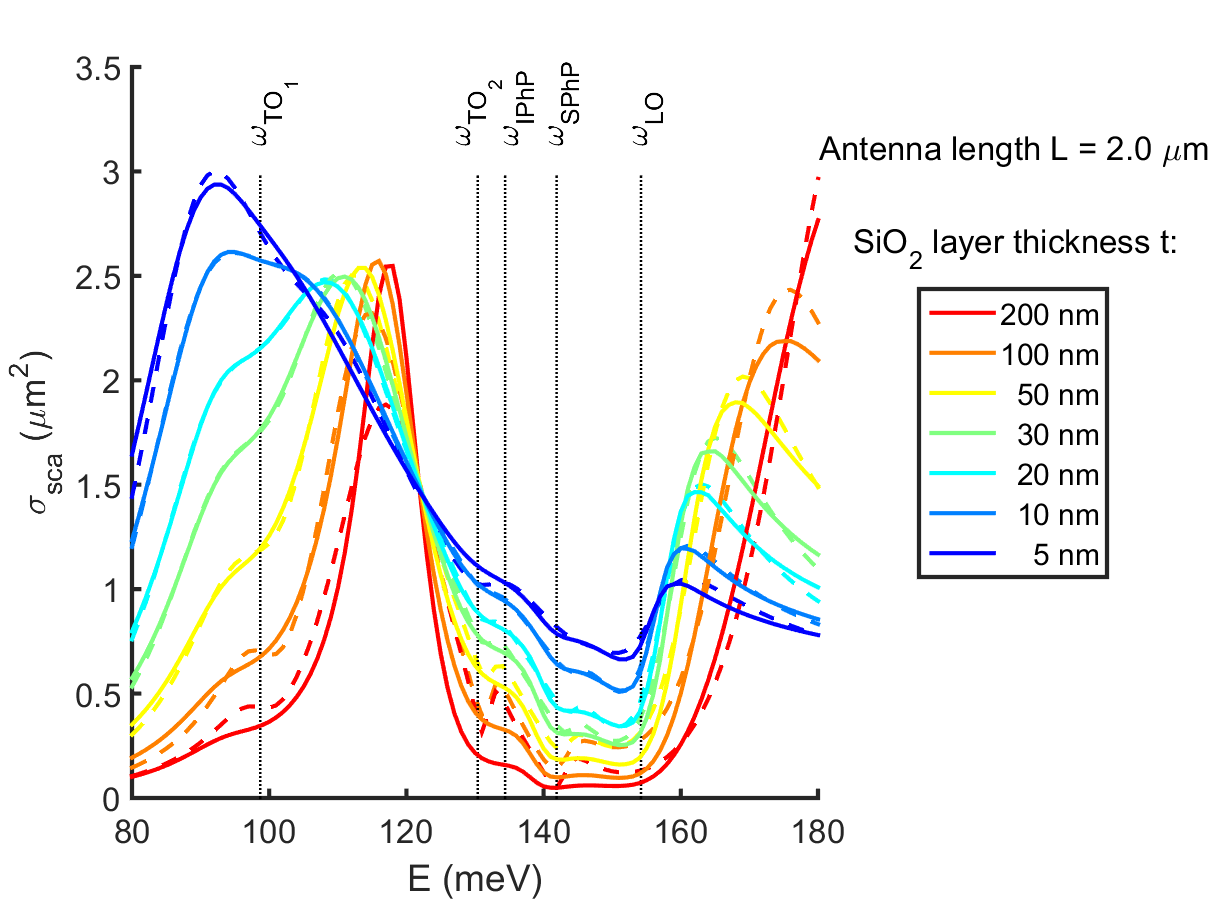}
\caption{Calculated (solid lines) and fitted (dashed lines) scattering cross section spectra of gold antennas with length $L=2.0\ \upmu$m on the SiO$_{2}$ layer for different thicknesses $t$ on the Si substrate.}
\label{fig_sigma_sca_L}
\end{figure}

\begin{figure}[h]
\includegraphics[width=\textwidth]{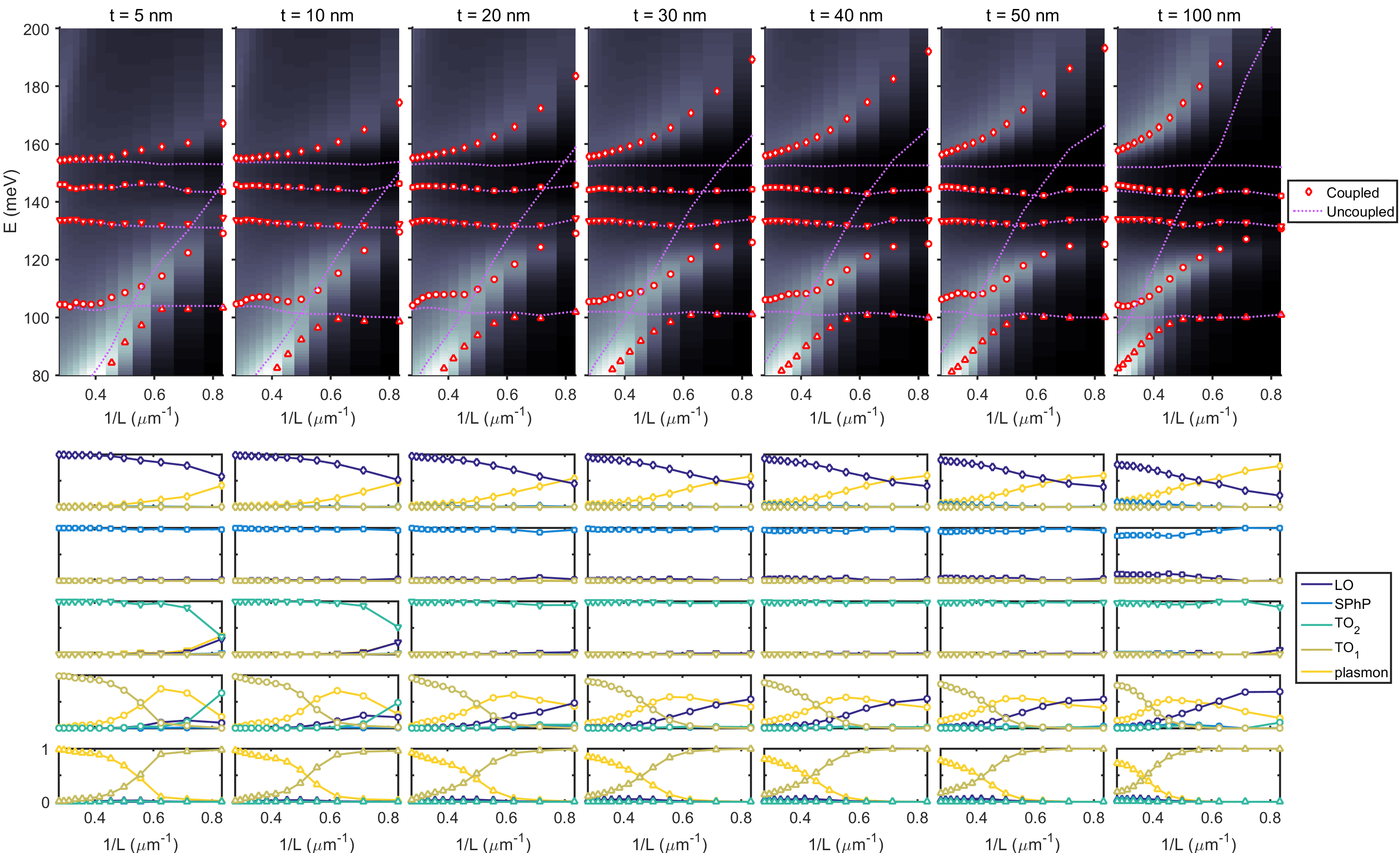}
\caption{(top) Dependence of the simulated scattering cross section spectra (background), uncoupled mode energies from fitting (magenta lines) and hybrid mode energies from subsequent analysis (red marks) on the inverse values of the antenna length $L$ for different SiO$_{2}$ layer thicknesses $t$. (bottom) Hopfield mixing coefficients showing relative contributions of uncoupled modes to the new hybrid modes. The bottom graphs correspond to the bottom branches in the images above (similarly for the other graphs and branches), the values for the Hopfield mixing coefficients are between 0 and 1.}
\label{fig_hopfield_SI}
\end{figure}